# Reduced leakage current in BiFeO$_3$ thin films with rectifying contacts

Yao Shuai,[1] Shengqiang Zhou,[1,2] Stephan Streit,[1] Helfried Reuther,[1] Danilo Bürger,[1] Stefan Slesazeck,[3] Thomas Mikolajick,[3] Manfred Helm,[1] and Heidemarie Schmidt[1]

*[1]Institute of Ion Beam Physics and Materials Research, Helmholtz-Zentrum Dresden-Rossendorf, P. O. Box 510119, Dresden 01314, Germany*

*[2]State Key Laboratory of Nuclear Physics and Technology, School of Physics, Peking University, Beijing 100871, China*

*[3]Namlab gGmbH, Nöthnitzer Strasse 64, 01187 Dresden, Germany*

**Abstract:** BiFeO$_3$ thin films were grown on Pt/c-sapphire substrates by pulsed laser deposition with different growth rates. With increasing growth rate the leakage current is decreased and the conduction mechanism changes from bulk-limited Poole-Frenkel emission to interface-limited Schottky emission. In the present work, we show that only the growth rate of the BiFeO$_3$ films close to the metal contacts has to be increased in order to reduce the leakage current and to observe saturated polarization-electric field hysteresis loops.

BiFeO$_3$ (BFO) is the most widely investigated single phase multiferroic material and has been the most promising candidate for multifunctional nonvolatile memory devices due to several advantages such as high Curie temperature (1123 K),[1] high Néel temperature (653 K),[1] and giant spontaneous polarization.[2] Nevertheless, its high leakage current density still remains a great challenge for practical applications, e.g. the ferroelectric random access memory.[3] Particularly, the leakage behavior of BFO thin films at high electric fields is deteriorated and has received much attention, because high electric fields are usually needed to switch the polarization of a BFO thin film due to its high coercive electric fields.

There have been many approaches to decrease the large leakage current in BFO thin films. Atomic substitutions at cationic A-sites or B-sites can reduce the leakage current of BFO thin films by 1~3 orders of magnitude due to the formation of defect complexes or the suppression of oxygen vacancies.[4-7] In addition, a solid solution approach where BFO and BaTiO$_3$[8] or BFO and PbTiO$_3$[9,10] are combined has also been reported to effectively decrease the leakage current. However, the ferroelectric properties of BFO are degraded if the doping or the solid solution approach was followed.

In the present work, pure BFO thin films without any dopant or solid solution compound were grown using fixed temperature and oxygen pressure (thermodynamic conditions), but different growth rates which were controlled by the laser repetition rate. As the growth rate increases, the leakage current of correspondingly prepared Au/BFO/Pt stacks is considerably suppressed, and the conduction at high electric field has been changed from bulk-limited to interface-limited mechanism. This transition is related with the modification of the interface between the electrodes and the thin film. Due to the low leakage current, the BFO thin film which is grown at higher rate shows a saturated polarization electric field (P-E) hysteresis loop.

BFO thin films were deposited on Pt (100 nm)/sapphire substrates by pulsed laser deposition (PLD). The thermodynamic parameters have been optimized to be 670 °C for the substrate temperature and 60 mTorr for the oxygen partial pressure. The laser fluence was kept at 2 J/cm$^2$, while the repetition rate was changed for different samples. The thickness of each film was measured by a step profilometer and the growth rate was calculated. The phase of the BFO thin films was detected by x-ray diffraction (XRD) using a Bruker D8 system with CuKα radiation. The surface morphology is characterized by atomic force microscopy (AFM) measurements. Auger electron spectroscopy (AES) has been used to determine the chemical composition of the stack. For electric property measurements, 0.3 mm$^2$ Au dots (top electrodes) were formed by magnetron sputtering. The leakage current was measured by a Keithley 2400. The sample labeling, growth condition and properties are listed in Tab. I.

Fig. 1 shows the XRD θ-2θ patterns of the BFO thin film samples. Besides the substrate, only BiFeO$_3$ has been identified in BFO-2Hz [Fig. 1(a)]. However, as the growth rate increases, an impurity phase is present in BFO-10Hz [Fig. 1(b)], which is believed to be Bi$_{25}$FeO$_{40}$ or Bi$_2$O$_3$. The same impurity phase has also been observed in BFO-10/2/10Hz [Fig. 1(c)]. Therefore, a high growth rate can result in the formation of a Bi-rich phase.

The influence of the growth rate on the stoichiometry of the thin films has been examined by AES. As shown in Fig. 2(a), BFO-2Hz exhibits a Bi-rich layer at the top interface between the Au electrode and the BFO thin film. By increasing the growth rate, the Bi content increases significantly in the Bi-rich layer [Fig. 2(b)], and the proportionate thickness of this Bi-rich is also extended. Furthermore, another Bi-rich layer at the bottom interface between the Pt electrode and the thin film is formed in BFO-10Hz with increasing growth rate, which can not be observed in BFO-2Hz. The AES results indicate that a higher growth rate aggravates the segregation of Bi near the interface. It is likely due to the fact that with high growth rate the deposition is at a supersaturated state, which favors the formation of the Bi-rich phase.[11] The segregation of a Bi-rich phase at the interface has been reported in other literatures.[12-13]

The leakage current, as shown in Fig. 3(a), has been suppressed by two orders of magnitude by simply increasing the growth rate. To get further insight into the electric properties, the conduction mechanism of the BFO thin films are investigated by plotting log$J/E$ vs. $E^{1/2}$ or log$J$ vs. $E^{1/2}$, corresponding to Poole-Frenkel emission (PF) or Schottky emission (SE)[14], respectively. PF emission is a bulk-limited, while SE is an interface-limited conduction mechanism. The following formulas describe the J-E characteristics related with these two conduction mechanisms:

$$J = BE \exp-(\frac{E_I}{kT} - \frac{q}{kT}\sqrt{\frac{qE}{\pi\varepsilon_0 K}})  \quad (1)$$

$$J = A^*T^2 \exp-(\frac{\varphi_b}{kT} - \frac{q}{kT}\sqrt{\frac{qE}{4\pi\varepsilon_0 K}})  \quad (2)$$

The J-E curve of the BFO-2Hz sample is well fitted to the log-linear form logJ/E~$E^{1/2}$ as illustrated in Fig. 3(b). The values of the optical dielectric constant K calculated from the slopes of the fitting using Eq. 1 are 5.68 and 6.51 for positive bias and negative bias, respectively. The reflection index n of pure $BiFeO_3$ has been reported to be 2.5.[15] Consequently an optical dielectric constant K=$n^2$=6.25 is expected, to which the calculated K values are very close. Therefore, PF is believed to dominate the conduction in BFO-2Hz.

On the other hand, the fitting for BFO-10Hz on a logJ/E~$E^{1/2}$ scale results in unreasonable optical dielectric constant K (not shown), which excludes the PF conduction mechanism. While the K values calculated using the fitting slopes in logJ~$E^{1/2}$ form and Eq. (2) amount to 6.3 and 7.3 for the two bias polarities, respectively [Fig. 3(c)]. These K values are in good agreement with the expected value of 6.25. With increasing growth rate, e.g. in the BFO-10Hz sample, the SE conduction mechanism dominates.

It has been reported that the Bi-rich phase in BFO thin films has a high conductivity and can cause a large leakage current[16,17]. Here we show that the BFO-10Hz which contains a Bi-rich layer at the interface has a considerably reduced leakage current. Therefore, the interface between the electrode and the thin films, favoring the interface-limited Schottky emission conduction, plays a more important role than the presence of impurity phase. In order to confirm this conclusion, we prepared the third sample BFO-10/2/10Hz. Although the leaky BFO-2Hz layer constitutes ca. 80% of the BFO stack in BFO-10/2/10Hz, the leakage current is still decreased by nearly two orders of magnitude as compared to BFO-2Hz, and is only slightly higher than that in BFO-10Hz [Fig. 2(a)]. As shown in Fig. 2(c), SE also dominates the high electric field region of BFO-10/2/10Hz, which is similar to BFO-10Hz. This is the direct evidence that the leakage current in the BFO-10Hz sample is mainly controlled by the interface and not by the bulk thin film if a high Schottky barrier is formed.

The growth rate does play an important role in BFO thin film growth, it has however been neglected in recent years. Only a few groups have paid attention to this key parameter. For example, Shelke *et al.*[18] grew epitaxial BiFeO$_3$ thin films on (100)-SrTiO$_3$ substrates with various repetition rates and observed significantly reduced leakage current in high repetition grown thin film. Fig. 1(d)~(f) illustrates the topography of the BFO thin films. With increasing repetition rate the corresponding surface roughness decreases from 14.1 nm to 4.2 nm. The low surface roughness enables the formation of a Schottky barrier between the top electrode and the thin film. Furthermore, a high growth rate can reduce island size and increase the island density during the very beginning of the growth,[19] leading to more compact thin films, which are also seen in Fig. 1(d)~(f). Both smooth surface and compact thin film significantly decrease the concentration of defects and surface charges at the top and bottom electrode interfaces. It is known that both Pt and Au possess high work function.

Therefore, in principle a Schottky contact should be formed when n-type BiFeO$_3$ thin films are in contact either with Au or Pt. However, the rough surface and a large amount of defects cause a large number of charges in BFO-2Hz and therefore considerably reduce the contact barrier.

Fig. 4 shows the P-E loops of BFO-2Hz, BFO-10Hz, and BFO-10/2/10Hz thin films. As expected, BFO-2Hz exhibits an unsaturated P-E loop with a remanent polarization Pr of only 27 μC/cm$^2$ due to the large leakage current. On the other hand, both BFO-10Hz and BFO-10/2/10Hz show well saturated P-E loops with nearly the same Pr of as large as 65 μC/cm$^2$, which is comparable with epitaxial BFO thin films grown on single crystalline substrate.[20,21]

In summary, BiFeO$_3$ thin films show suppressed leakage currents as the growth rate increases. A high growth rate induces a higher Schottky barrier at the interface between the electrode and thin film and thus suppresses the leakage current particularly at high electric field and enables saturated polarization electric field hysteresis loop. Our work can help to guide future work on ferroelectric memory devices where ferroelectric thin films with high resistance and low leakage current have to be grown on a metal bottom.

Y.S. would like to thank the China Scholarship Council (grant number: 2009607011). S.Z., D.B., and H.S. thank the financial support from the Bundesministerium für Bildung und Forschung (BMBF grant number: 13N10144).

Tab. I. Sample identification and description of growth parameters and conduction mechanisms

| Sample | Repetition rate (Hz) | Growth rate (nm/s) | Thickness (nm) | Conduction mechanism |
|---|---|---|---|---|
| BFO-2Hz | 2 | 0.04 | 312 | Poole-Frenkel |
| BFO-10/2/10Hz | 10/2/10Hz | — | 326 | Schottky Emission |
| BFO-10Hz | 10 | 0.21 | 358 | Schottky Emission |

**FIGURE CAPTIONS**

Fig. 1. XRD pattern of (a) BFO-2Hz, (b) BFO-10Hz, and (c) BFO-10/2/10Hz, and surface topography of (d) BFO-2Hz, (e) BFO-10Hz, and (f) BFO-10/2/10Hz, respectively.

Fig. 2. Depth profile of the (a) BFO-2Hz and (b) BFO- 10Hz measured by Auger electron spectroscopy.

Fig. 3. J-E curves of BFO thin films plotted on (a) $\log J \sim E$, (b) $\log J/E \sim E^{1/2}$, (c) $\log J \sim E^{1/2}$, and (d) $\log J/E^2 \sim 1/E$ scale.

Fig. 4. Polarization-Electric field loops of BFO-10Hz, BFO-10/2/10Hz, and BFO-2Hz.

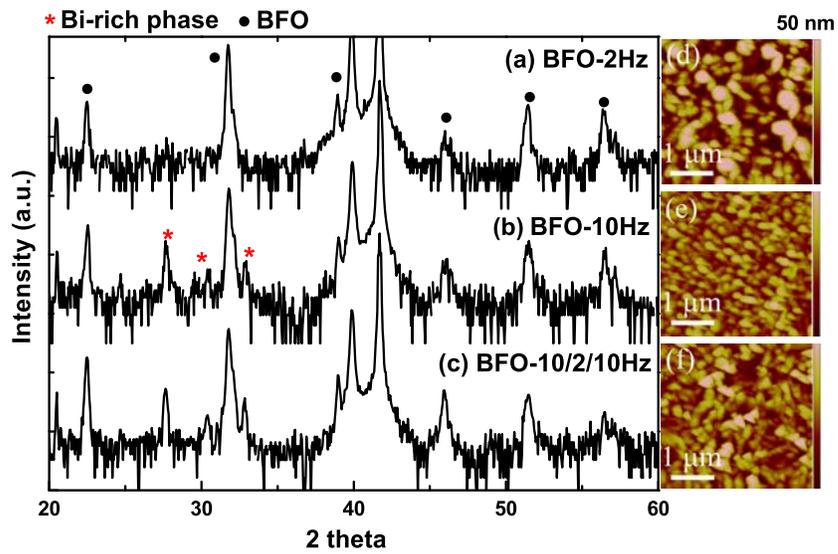

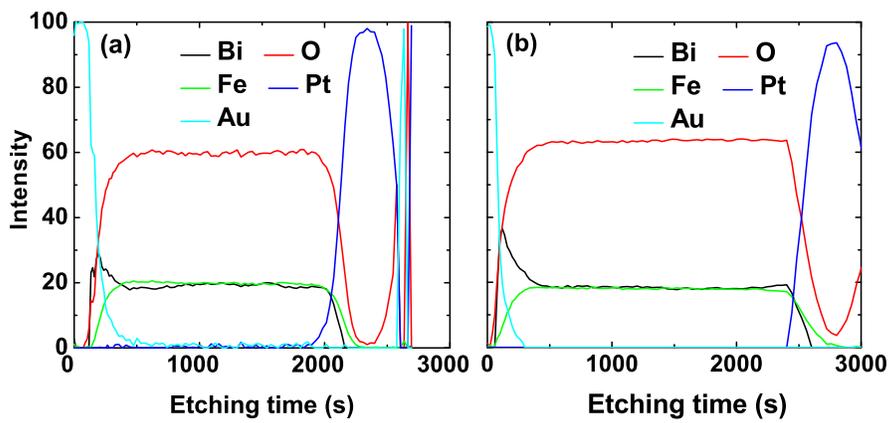

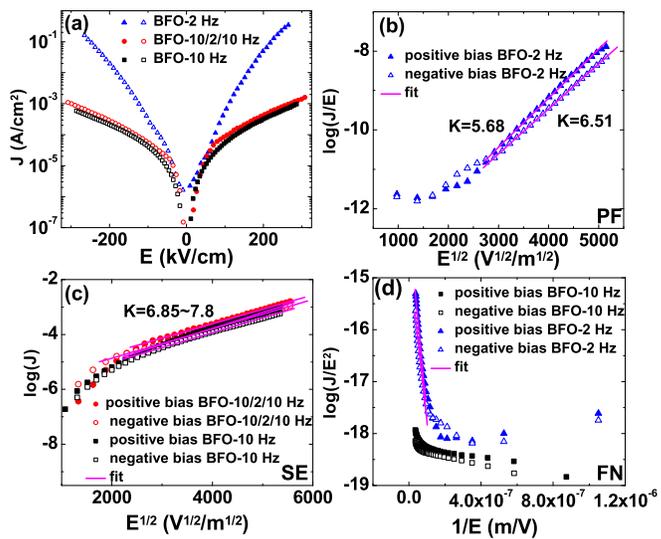

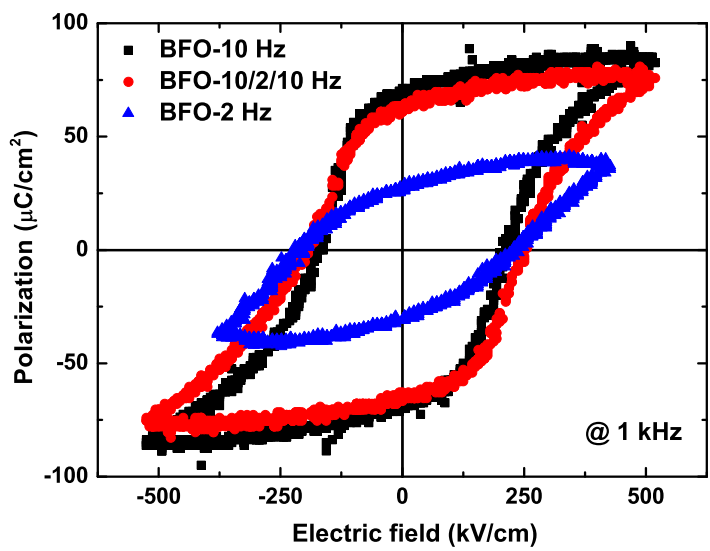